# New telescope designs suitable for massively-multiplexed spectroscopy

Luca Pasquini[1a], B. Delabre[a], R. Ellis[a], T. de Zeeuw[a,b]

[a]ESO, Karl Schwarzschild Strasse 2, D-85748 Garching b. München, Germany
[b]Leiden Observatory, Leiden University, Niels Bohrweg 2, 2333 CA Leiden, Netherlands

## ABSTRACT

We present two novel designs for a telescope suitable for massively-multiplexed spectroscopy. The first is a very wide field Cassegrain telescope optimised for fibre feeding. It provides a Field Of View (FOV) of 2.5 degrees diameter with a 10m primary mirror. It is telecentric and works at F/3, optimal for fibre injection. As an option, a gravity invariant focus for the central 10 arc-minutes can be added, to host, for instance, a giant integral field unit (IFU). It has acceptable performance in the 360-1300 nm wavelength range. The second concept is an innovative five mirror telescope design based on a Three Mirror Anastigmatic (TMA) concept. The design provides a large FOV in a convenient, gravity-invariant focal plane, and is scalable to a range of telescope diameters. As specific example, we present a 10m telescope with a 1.5 degree diameter FOV and a relay system that allows simultaneous spectroscopy with 10,000 mini-IFUs over a square degree, or, alternatively a 17.5 square arcminutes giant IFU, by using 240 MUSE-type spectrographs. We stress the importance of developing the telescope and instrument designs for both cases.

## 1. INTRODUCTION

Many branches of astrophysics will be transformed in the coming years by the ongoing and planned large photometric surveys, such as GAIA, Euclid, DES, HSC, Javalambre, CRTS, PTF, ASAS-SN, HITS, CHASE, Pan-STARRS and the VISTA and VST public surveys. The Large Synoptic Survey Telescope (LSST) will eventually dwarf all of these. It is therefore clear that the next big quest will be for spectroscopic surveys, matching as much as possible the photometric ones, as already anticipated in the MSE science cases[1] and in several decadal reports. Presently several large spectroscopic surveys are in progress (APOGEE, GAIA-ESO), and new wide field instruments are being built for existing telescopes, including MOONS at the VLT[2], 4MOST at VISTA[3], PFS at Subaru[4] and WEAVE at WHT[5]. The overall characteristics of these new facilities can be summarized as telescopes having at most A×FOV ~ 60 (A is the area of the primary in square meters, FOV in square degrees) and $N_{obj}$ ~1000-5000 fibres. They each plan surveys of several tens of millions of objects over several thousands of square degrees in the sky.

How could a new seeing-limited telescope advance the subject with respect to these upcoming facilities? Adopting a simple parametric view, one would like to increase the number $N_{obj}$ as well as A×FOV. These two parameters will directly affect the speed and reach of the surveys. Ideally, A×$N_{obj}$×FOV should be one order of magnitude larger than for the currently planned facilities. In addition, one would like higher quality spectra: this could be expressed in different ways, for instance: higher fidelity (better sky subtraction, flux calibration) and/or higher resolution (spectral and spatial). Furthermore, an increased discovery space might follow avoiding object pre-selection, such as is possible in a giant integral field unit (IFU). An increased versatility would follow access to a larger spectral range, or by providing several (simultaneous) observing options. For this exercise it is essential to conceive the telescope and its associated instrumentation as a single system; we imagine such instrumentation will represent a considerable fraction of the overall cost of the facility. We have therefore investigated two design options following very simple requirements, not driven at the moment by specific science cases, but rather by the basic considerations mentioned above. Our study parallels a scientific investigation of the merits of such a facility being conducted by the ESO community[6].

One option focuses on extending the existing capabilities, enhancing and optimizing the first two parameters: $N_{obj}$, but mostly the FOV. This concept requires in addition that the telescope is fibre friendly, working at an appropriate F aperture (~F/3), with an atmospheric dispersion corrector (ADC) and an easily accessible telecentric focal plane.

---

[1] Email:lpasquin@eso.org

The second design complies with the very stringent requirements of having a large FOV and being very versatile, by hosting more than $10^4$ mini IFUs or a giant IFU. No ADC is required, but the instrument must be located in a gravity invariant focus.

## 2. THE FIBRE TELESCOPE

The design presented here is a Cassegrain telescope, with a three-lens corrector (four aspheric surfaces). Following a concept introduced for other correctors[7,8], by laterally moving independently the corrector and the first lens, atmospheric dispersion can be compensated. The secondary is also an aspheric, 3.7m in diameter; this is large but smaller than the secondary of the E-ELT. The optical design is shown in Figure 1, the corrector in Figure 2. It offers for a 10m telescope at a convenient F/3 (ideal for fibre injection): a FOV of 2.5 degrees diameter, or 4.9 square degrees, 3.5 times that of the Subaru PFS; the same as VISTA, but for a telescope with a 2.5 times larger aperture. The convenient plate scale of 145 µm/arcsecond provides a focal plane diameter as large as 1.3 meter, that is about 2.5 times larger than Subaru and VISTA. 10,000 fibres or more could therefore be hosted in the focal plane using the same technology developed for these facilities. The telescope design is very compact and cost-effective; the distance between primary and secondary is 6.8m and there is no Nasmyth platform in this simple configuration. The fibre system and spectrographs have not been designed yet, but no major challenges are envisaged. The detailed design will depend on the precise requirements, but a major advantage of fibres is the de-coupling between focal plane pick–up and spectrographs. Still ~50 spectrographs will be needed, if for instance one would clone those for 4MOST. Figure 3 shows the chromatic image quality (360-1300 nm) over the 2.5 degrees diameter field at different airmasses, with the working ADC. The focal plane location is very convenient for maintenance and minimizes the fibre length, when compared, for instance, to a prime focus design.

By inserting two mirrors in the optical path, it is possible to create a 10 arc-min diameter central FOV located at a convenient gravity-invariant Coudé focus, to host a giant IFU, for instance a larger version of MUSE[9], as shown in Figure 4. The Cassegrain corrector does not need to be removed for this operation: M5 and M6 are permanently mounted. .

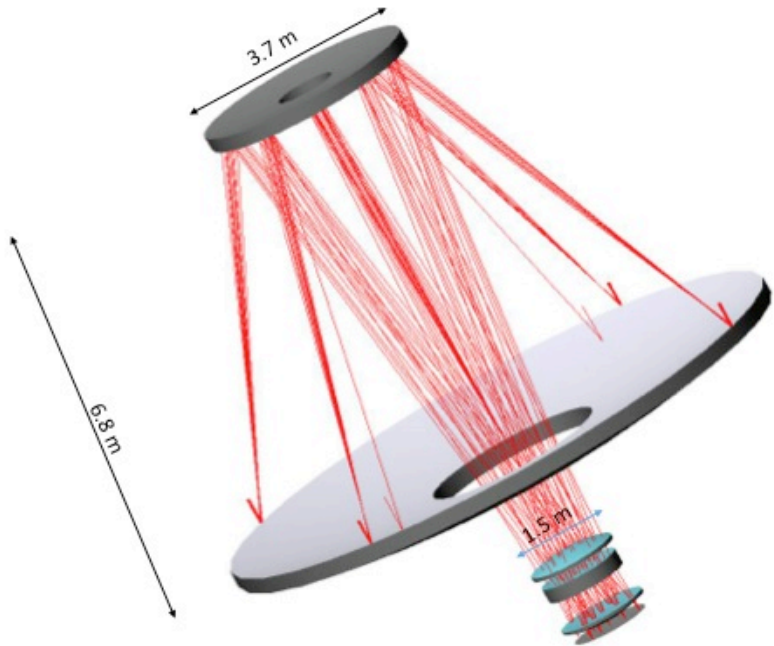

**Figure 1: The 10m Cassegrain Fibre Telescope. The corrected FOV is 2.5 degrees diameter. The FOV is limited by the maximum size of available corrector lenses (1.5m).**

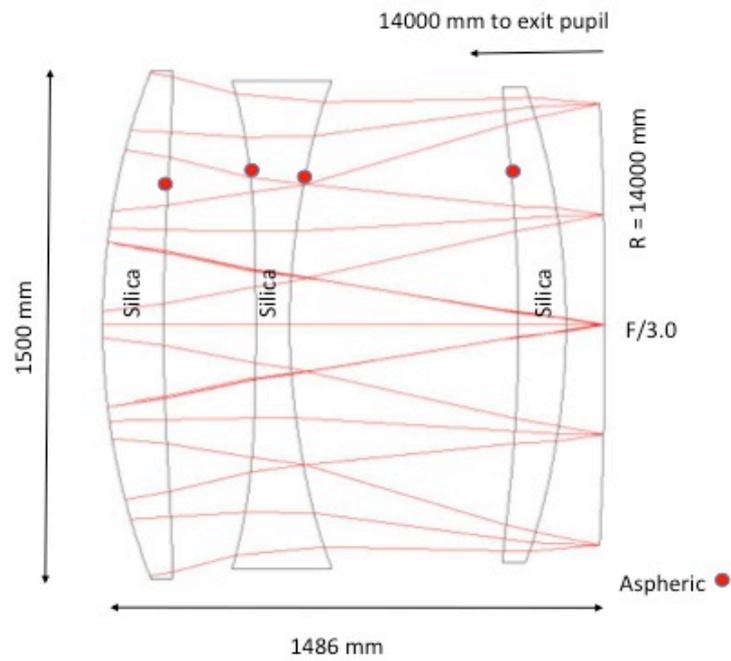

Figure 2: The corrector. Surfaces with red points are aspheric. By shifting the whole corrector and the front lens, ADC correction is obtained.

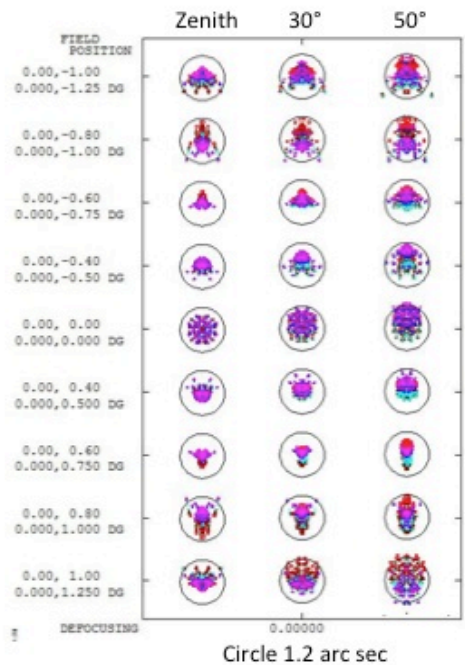

Figure 3: Polychromatic image quality (360-1300 nm) at different locations in the FOV and at different zenith distances.

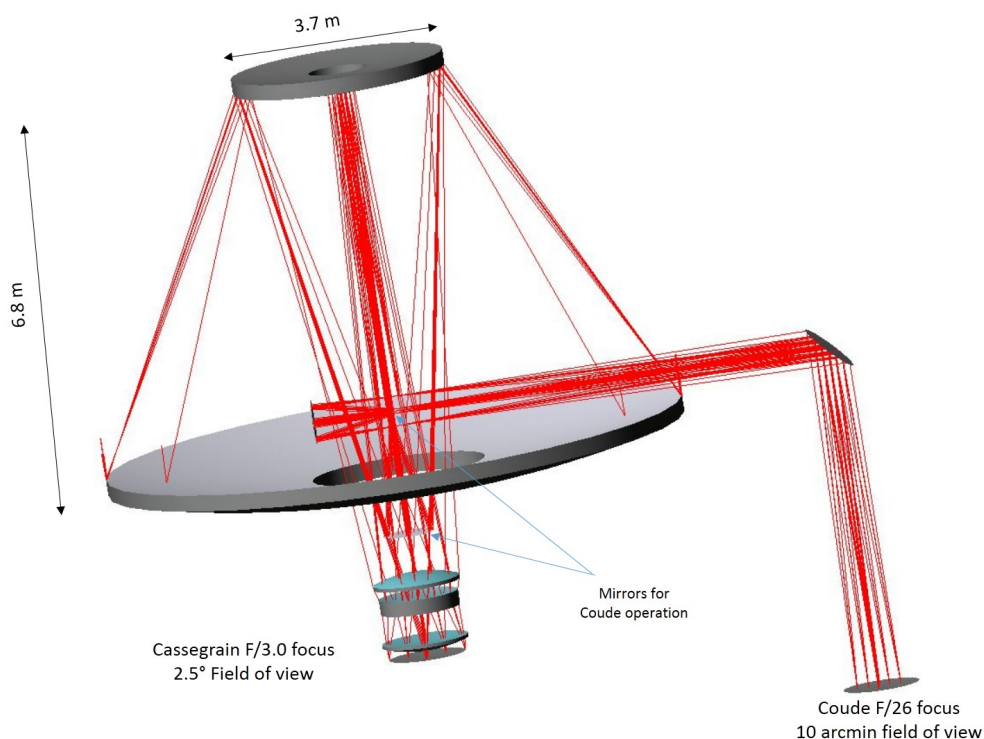

**Figure 4: By inserting two mirrors (M3 and M4) in front of the corrector, it is possible to activate a 10 arcmin central Coudé focus, gravity invariant, suitable to host a giant IFU. M5 and M6 will be permanently mounted.**

## 3. THE RING FIELD TELESCOPE

Although the concept of a three mirror anastigmatic (TMA) telescope is certainly not new, the combination of its large FOV, large gravity-invariant focal plane, and versatility makes this telescope extremely instrument-friendly for the present purpose. It can also be scaled to any diameter. Specifically, for this case, the central 20 arcminute field, not used by the main (Coudé) focus, can host an additional instrument (for instance pick-up fibres for a high resolution spectrograph) at Nasmyth, with the advantage that both foci (Coudé and Nasmyth) can be used simultaneously. In addition, at the main focal plane several instruments could be used simultaneously, given the rotating focal plane is very large.

### 3.1 The telescope

Our telescope design is based on a TMA design, M1, M2 and M4 are aspheric and the center of the FOV is not available to the main (Coudé) focal station. The distance between primary and secondary is less than 8.6m. M4 is large, but with acceptable manufacturing dimensions (3.5m). The whole Coudé platform must rotate to compensate the field rotation, but no adapter/rotator is needed. The last mirror, M5, is in the pupil and can be used to compensate for run-out errors of the platform and for tip-tilt correction. At F/17.7 it provides a plate scale of 857 microns/arcsecond. The corrected 1.5 degree diameter FOV is therefore very large (4.6m diameter) and can comfortably pick up the light for several instruments. The telescope design is shown in Figure 5. The instruments are in a gravity invariant location. This design does not require an ADC, although a large and therefore segmented linear ADC could be inserted between M3 and M4.

In principle, the 20 central arcminutes (i.e. almost as large as the Nasmyth focus of the VLT) can be made available by inserting a mirror back to back of M5 (thus avoiding vignetting the main focus), and then, through a three mirror system, brought to a F/15 Nasmyth location, as shown in Figure 6. M6 is near to the pupil and could be used also as deformable mirror to enable AO corrections. For this telescope, it is worth noticing that the two foci are not mutually exclusive, so could be used for simultaneous observations. Figure 6 shows the design of the telescope with both foci enabled. The image quality is excellent, both at the Coudé as well as at the Nasmyth, as shown in Figure 7.

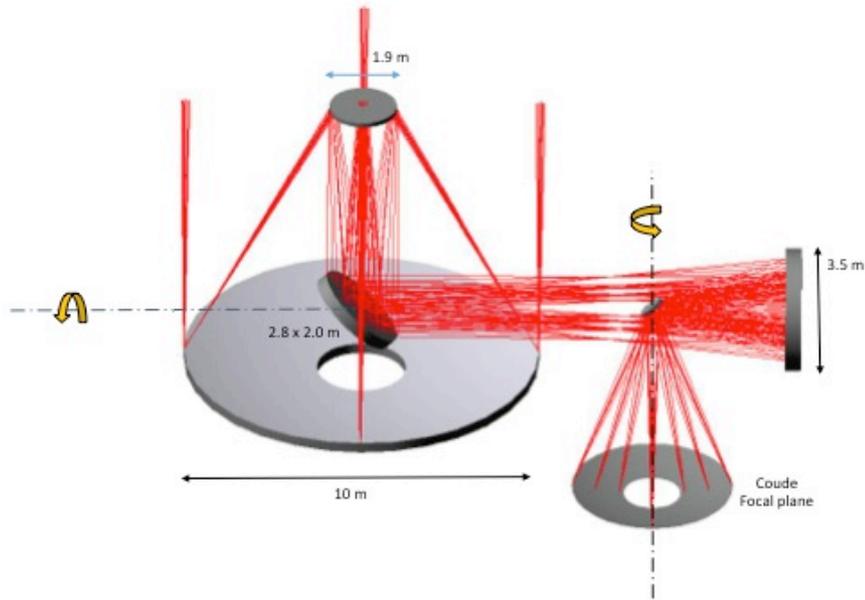

Figure 5. The 5-mirrors 10m TMA design. It provides a FOV of 1.5 degrees in diameter, with a 30 arcmin central obscuration, at a gravity invariant Coudé focus. Several instruments can be hosted and in principle used simultaneously. The design has no ADC, but a linear ADC could be added between M3 and M4.

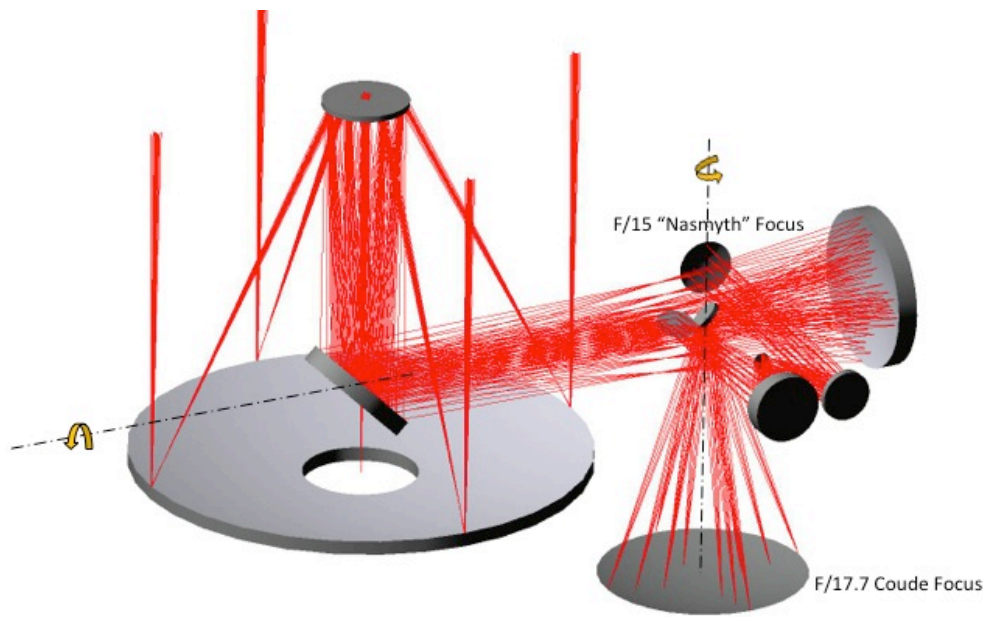

Figure 6: The TMA telescope with the two foci, the Coudé and the 20 arcmin Nasmyth, simultaneously enabled.

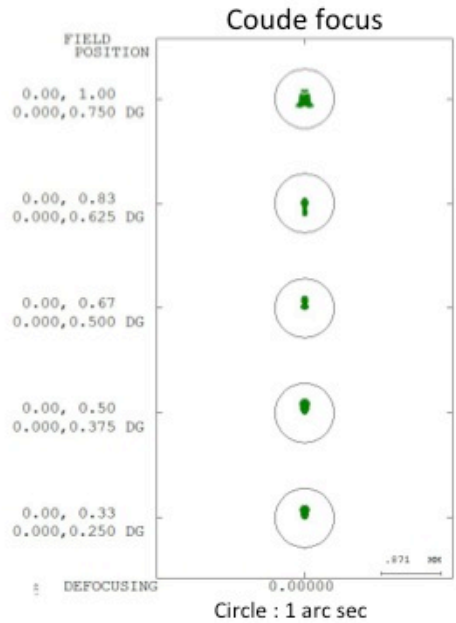

**Figure 7: Image quality at the Coudé focus. The circles are 1 arcsecond in diameter**

### 3.2 The pick object, relay optics and spectrograph

One challenge for any MOS telescope (not using fibres) is the transmission of the light from the focal plane to the spectrograph slit. We therefore considered it essential to study the relay optics systems together with the telescope. We have linked it to a MUSE – type image slicer and spectrograph, to produce an end-to-end system. In order to conceive the relay system, we have assumed the use of mini-IFUs. The reasons for this choice are both technical and scientific. Technically the mini – IFU solution is the most challenging (and does not require an ADC). Fibres, for instance, would be rather trivial in this location (it is a configuration very similar to e.g. FLAMES at the VLT[10]) and only the large number of objects would represent a challenge. Scientifically the use of true IFUs (opposed for instance to fibres IFUs) has been proven to produce very high quality data, for instance with MUSE and KMOS[11] at the VLT. They offer spatial resolution, optimal sky subtraction and enhanced sensitivity (no aperture flux losses).

**Pick Object**

Half of the focal plane is covered by stripes of mirrors. Each mirror determines the patrol area of a single IFU. These mirrors are scanned by the object selection mirrors, located in the pupil. By changing the angle of these object selector mirrors the focal plane is patrolled, and the light is then re-organized into a series of 2.5x2.5 arcseconds stamps along slits. We have chosen mini IFUs of 2.5x2.5 arcseconds in size, but this size can be changed. The size of the focal plane mirrors determines the patrol area. In the configuration shown in Figure 8, for instance, each mini-IFU patrols a 12x12 arcsecond area, so it will acquire the spectrum of one 2.5x2.5 arcsecond portion of the sky in each patrolled area. The mirrors are organized in stripes to allow the transmission of the light by the object selector mirrors.

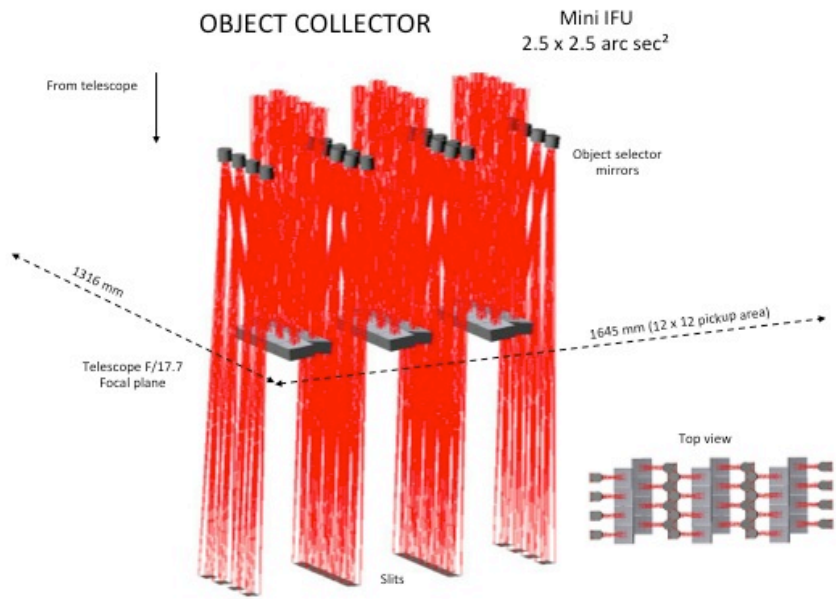

**Figure 8: The pick object configuration. Focal plane is filled with stripes of patrol mirrors and the object selector mirrors select a 2.5x2.5 arceconds area within a patrol area (12×12 arcesecond in this figure).**

### Relay Optics & Spectrograph

After the pick object, an anamorphic optical system relays the light and organizes the mini-IFUs for ingestion into the MUSE-type image slicer and spectrograph. An end-to-end picture of the system is given in Figure 9.

The MUSE design has been modified to accept an 8×4K detector, in order to maximize the number of objects. By keeping the original MUSE sampling (0.2×0.1 arcseconds), 42 2.5x2.5 arcseconds mini-IFUs are accepted by one spectrograph. With 240 spectrographs it is therefore possible to observe 10,000 objects simultaneously.

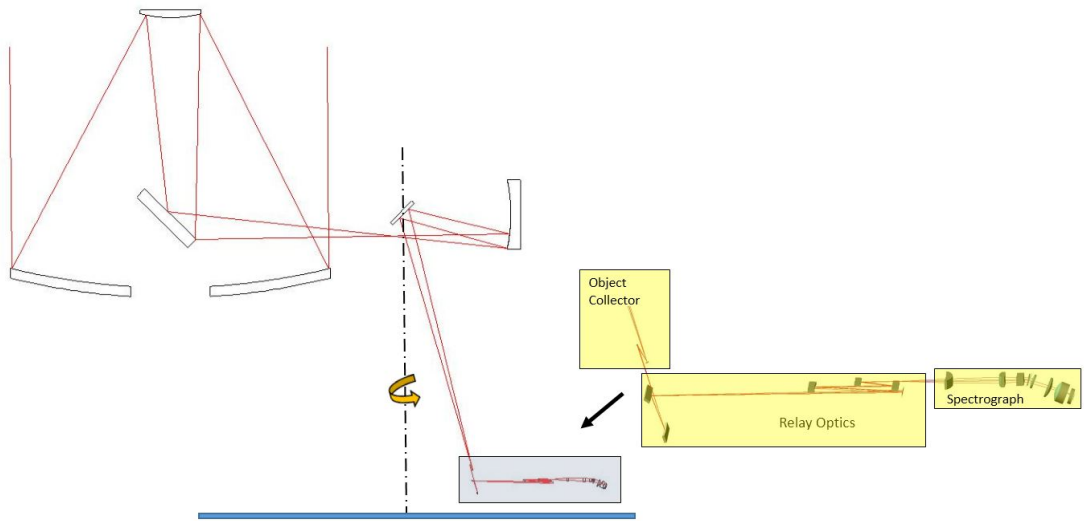

**Figure 9: The end-to-end concept for the TMA telescope, from sky to detector. It includes the pick object, relay optics, slicer and modified MUSE-type spectrograph.**

**Some performance estimates and a giant IFU**

As a reference, MUSE at the 8m VLT covers the 465-930 nm spectral range, and produces at R=3000 a S/N=9 spectrum for a R(Vega) = 22 magnitude object in 30 minutes. With patrol areas of 24 arcseconds, a FOV of 0.9 square degrees would be covered with 2 integrations, providing 20000 high quality spectra. Of course another patrol area or mini-IFU size can be chosen for the final design.

It is possible to use a similar arrangement to observe a contiguous area equivalent to a giant IFU. Using the same concept, however, it is possible only to observe stripes within a contiguous area. Two observations, shifted by a few arcseconds, would be needed to obtain a full 3D area coverage. The sky area covered in this way, with 240 spectrographs and 2 exposures, will be 35 square arcminutes.

We recall that the fine MUSE sampling (0.1×0.2 arcseconds) at the VLT is required by the use of this instrument with GLAO. In principle, a coarser sampling can be acceptable in a seeing limited case. This would make the camera optics more challenging but at the same time allow either a larger number of objects/spectrograph, or a larger mini-IFU size, or a larger area monitored by the giant IFU. No trade-off study has been performed yet.

## 4. DIAMETER, COSTS, DEVELOPMENT

As mentioned in the introduction, since the instrument in such a facility represents a non negligible fraction of the costs, cost estimates should include both the telescope and spectrographs. The telescope costs can be roughly estimated at ~130 M€ (without the Focal Plane system). This value applies either to an 8m monolithic (50m$^2$ M1 area) or to a 10m class segmented (~75m$^2$ M1 area) and it is simply computed by analogy with available cost figures (corrected for inflation) for VLT, GEMINI, Keck and GTC. There are three factors that may help in containing the costs:

- The use of an existing site, such as Cerro Montura (Paranal) or La Silla, will minimize the infrastructure costs. The costs of some of the previous telescopes include in fact infrastructure development.
- Synergy with E-ELT, in particular the re-use of the E-ELT M1 segments and support. These segments and their supports are light (370 Kg for a 1.45m mirror, 1.36 m$^2$), and could be re-used without need of study, design and prototyping; only the mirrors figures should be changed.
- The spectroscopic telescope is compact and may have only one focus: weight, dome size are reduced, so it may reduce two cost drivers simultaneously: low M1 weight and small dome and main structure.

The instrument cost is expected to be of the same order of the telescope. Assuming for instance 10 MUSEs (240 spectrographs), their estimated cost (including capital costs and resources) would be ~130 M€. Of course, for such a massive production the costs will critically depend on the optimization of the production process. The work cannot be organized in the same fashion as for typical astronomical instruments, but must be automatized. It seems therefore reasonable to assume for an 8m monolithic or a 75 m$^2$ segmented a total order of magnitude costs of 260 M€ with an approximate 50:50 split between telescope and instruments, including staff effort. The associated uncertainty is about ±25% and there are good reasons to believe it can be done cheaper.

### 4.1 Scaling the Costs with Diameter:

The classical formula for telescope costs scale as ~ $D^{2.7}$. However, with the hypothesis of maintaining the same instrument, the costs depend less strongly on D. With instrument and telescope split equally for a 8m monolithic or a 10m segmented $C/C_{8m} = ((D/8)^{2.7} + 1)/2$. For a segmented telescope, D may not be the best figure, and we use instead square root of the M1 area A. $C/C_{10m} = ((A/73.5)^{1.35} + 1)/2$. The results are shown in the table below.

| Diameter | N Tiles | Area (m²) | Cost/8m | Cost (M€) |
|---|---|---|---|---|
| 8.8 | 30 | 40.8 | 0.73 | 190 |
| 11.3 | 54 | 73.4 | 1 | 260 |
| 13.9 | 84 | 114.2 | 1.4 | 364 |

**Table 1 Expected costs the different telescope diameters. A is the primary area, and M1 is assumed to be segmented. Tiles have the dimensions of the E-ELT ones.**

A is the area of the primary assuming that the central 7 segments have been taken out for the central hole. The 8-14 meters range with a complete spectrograph (10 MUSE) is therefore feasible within a factor two overall cost range.

The D=14m solution is rather interesting. A substantial gain with respect to the existing telescope area is much desirable, especially for the mega-IFU case, but it will come at expense of the size of the FOV. The 14m option would imply a factor two (or more) gain in area with respect to existing telescopes providing, together with the 5 times more objects, one order of magnitude gain.

By working with a telescope of the 10 – 14m diameter class, one problem is the matching of the pixel size. For instance, for a 14m telescope, an F/2 camera provides a pixel sampling of 0.12 arcsec/pixel. This is sub-optimal and, as already occurs for Subaru PFS and MOONS, fast cameras (close to F/1) must be developed. When considering that, for a spectroscopic facility, tens or hundreds of spectrographs are needed, the most cost effective investment is to develop a cheap, fast camera. With a 14m telescope, an F/1.1 camera provides a sampling of ~0.23 arcsec/pixel (15 μm pitch), that is a very interesting value. The design proposed in Figure 9 is extremely efficient and simple, but requires a curved detector. Curved devices seem to have a rather wide range of applications beyond astronomy, so their availability in a short time may be practical.

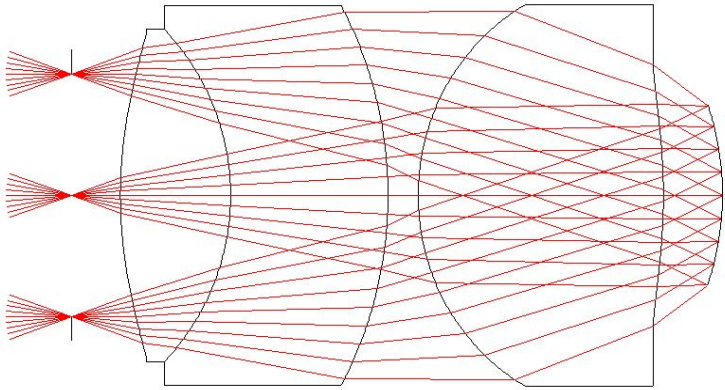

**Figure 10: A F/1.1 Camera offers a superb image quality, and very high transmission, with a 10 cm beam that feeds a 4kx4k 15 μm curved detector. The first and the last two surfaces are aspheres.**

## Preparatory R&D

It is worth recalling that the speed of a survey is, in the first instance, fixed by the spectrograph number and camera aperture, independent of the telescope diameter D. For a seeing-limited telescope, the flux $N_{phot}$ scales as $D^2$ but the pixel sampling scales with the same law, therefore the flux/pixel is independent of D. Similarly, as the FoV scales as $D^{-2}$, for a given area and number of objects, the survey is independent of the telescope diameter. Only for a giant IFUs is this not the case because the number of objects increases more steeply with the limiting flux than with the surveyed area. Therefore a large telescope observes more objects than a small one for the same exposure time, in spite of the smaller field.

When considering this, it is clear that the larger D and FOV can only be fully exploited by improving $N_{obj}$. This implies that the investment in a larger telescope D must be balanced by a proper investment in the number of spectrographs (scales as $D^2$), or that fast, cheap cameras must be produced, to optimize the sampling.

Considering the potential challenges for such a facility, several areas of R&D can be envisaged driven by considerations of the mass production of spectrographs and challenges of the focal plane pick and relay system:
- Fast, cheap cameras & cost-effective spectrograph design
- curved CCDs, controllers and cryogenics
- automation of spectrograph alignment
- focal plane split & relay system

## 5. CONCLUSIONS

In response to the growing demand for a transformational massively-multiplexed spectroscopy facility, we have studied several telescope design options,. We present two rather different solutions that would provide a quantum leap in capability with respect to the facilities presently under construction. The first is ideal for fibres. With a 2.5 degree diameter Cassegrain FOV on a 10 meter telescope, it enhances by one order of magnitude the survey efficiency of planned facilities. The telescope, in addition, can easily provide an additional gravity-invariant focus. The second concept provides a 1.5 degree FOV, corrected, huge Coudé focus. It is extremely versatile and ideal for gravity invariant instruments. The central 20 arcminutes can be diverted to a Nasmyth focus for simultaneous observations. For this telescope, an end-to-end concept has been designed, suitable for many mini-IFUs or a giant one. We emphasize that the cost of the telescope can highly benefit from the on-going developments for the E-ELT and from the location in an existing infrastructure, and we also find that the cost of the instruments are comparable with that of the telescope. Finally, we highlight the relevance of developing fast, cheap cameras and we identify some priority areas, such as curved detectors, for relevant R&D.

## 6. ACKNOWLEDGEMENTS

The authors thank R. Bacon, P. Callier, M. Quattri, J. Spyromilio for very interesting discussions and valuable input.